\documentclass{article}

    \PassOptionsToPackage{numbers, compress}{natbib}

\usepackage[main, final]{neurips_2026}

\usepackage[utf8]{inputenc} 
\usepackage[T1]{fontenc}    
\usepackage{hyperref}       
\usepackage{url}            
\usepackage{booktabs}       
\usepackage{amsfonts}       
\usepackage{nicefrac}       
\usepackage{microtype}      
\usepackage{xcolor}         

\usepackage{amsmath} 
\usepackage{amsthm} 
\usepackage{amssymb}	
\usepackage{multirow}
\usepackage{booktabs}
\usepackage{tabularx}
\usepackage{cleveref}
\usepackage{mathtools}
\newtheorem{theorem}{Theorem}[section]

\newtheorem{corollary}[theorem]{Corollary}

\newtheorem{definition}{Definition}[section]

\DeclarePairedDelimiter\rbra{\lparen}{\rparen}

\DeclarePairedDelimiter\abs{\lvert}{\rvert}
\DeclarePairedDelimiter\Abs{\lVert}{\rVert}

\DeclarePairedDelimiter\ket{\lvert}{\rangle}
\DeclarePairedDelimiter\bra{\langle}{\rvert}

\newcommand{\braket}[2]{\left< #1 \vphantom{#2} \middle| #2 \vphantom{#1} \right>} 
\newcommand{\ketbra}[2]{\ensuremath{\ket{#1}\bra{#2}}}
\newcommand{\tr} {\operatorname{tr}}
\renewcommand{\epsilon}{\varepsilon}

\title{Quantum Composite Hypothesis Testing with Small Error}

\author{%
  Chenghua Liu \\
  Institute of Software, Chinese Academy of Sciences \\
  University of Chinese Academy of Sciences \\
  Beijing, China \\
  \texttt{liuch.russell@gmail.com}
  \And
  Qisheng Wang \\
  School of Computer Science \\
  Shanghai Jiao Tong University \\
  Shanghai, China \\
  \texttt{QishengWang1994@gmail.com}
}

\begin{document}

\maketitle

\begin{abstract}
 We present a method of quantum composite hypothesis testing with small error,
  which enables us to establish quantum lower bounds in nonparametric statistics and 
  high-dimensional functional estimation. 
    We extend the phase estimation lower-bound argument of \hyperlink{cite.MdW23}{Mande and de Wolf (ESA 2023)}
    to composite hypotheses for one-parameter unitary families. 
    As an application, we settle the quantum complexities for a series of problems in 
    property testing and functional estimation with small error:
    \begin{itemize}
        \item For $\ell_2$-closeness testing, we show that the approach due to \hyperlink{cite.LWL24}{Luo et al.\ (\textit{IEEE Trans.\ Inf.\ Theory} 2024)} is optimal.
        \item For Tsallis entropy estimation, where $q = 2$ refers to the Gini impurity, 
        we show that the approaches due to \hyperlink{cite.BCWdW01}{Buhrman et al.\ (\textit{Phys.\ Rev.\ Lett.}\ 2001)} 
        and \hyperlink{cite.EAO+02}{Ekert et al.\ (\textit{Phys.\ Rev.\ Lett.}\ 2002)} are optimal 
        for integer $q \geq 2$ and the approach due to \hyperlink{cite.CGLWW26}{Chen et al.\ (ICALP 2026)} is near-optimal for real $q \geq 1.5$. 
        \item For pure-state Uhlmann fidelity and trace distance estimation, we show that the 
        approach due to \hyperlink{cite.Wan24}{Wang (\textit{IEEE Trans.\ Inf.\ Theory} 2024)} is optimal. 
    \end{itemize}

\end{abstract}

\section{Introduction}
Hypothesis testing is a fundamental problem in statistics.
A testing problem is specified by two mutually exclusive hypotheses: a null hypothesis \(H_0\) and an alternative hypothesis \(H_1\).
Given observations generated under one of the two hypotheses, the goal is to decide which hypothesis is true with a prescribed error probability.
Characterizing the complexity of such testing problems is central to information theory, statistics, and learning theory.
One of the classical lower-bound techniques is Le Cam's two-point method \cite{LC73,Yu97}, where both hypotheses are point hypotheses.
This idea has been extended to composite hypotheses, for example through the method of two fuzzy hypotheses \cite{Tsy09}.
Such techniques have played an important role in proving minimax lower bounds for nonparametric statistics and functional estimation; see, for example, \cite{Tsy09,JHW15}.

Quantum computing has brought significant speedups to property testing problems through
its ability to concentrate probability mass more efficiently
(e.g., via quantum amplitude estimation \cite{BHMT02}). 
A growing line of work has developed quantum algorithms for distributional property testing and related functional-estimation tasks \cite{BHH11,CFMdW10,Bel19,GL20,LW19,WZL24,LWL24}.
However, the quantum complexity of many fundamental testing problems remains open, especially in the small error regime.
A notable feature of the quantum setting is that the dependence on the error probability \(\delta\) varies substantially across problems.
For example, (i) logarithmic dependence: quantum phase estimation to within additive error 
$\epsilon$ with small error $\delta$ requires query complexity 
$\Theta\rbra{\log\rbra{1/\delta}/\epsilon}$ \cite{MdW23}; 
(ii) sub-logarithmic dependence: quantum search over $n$ elements with small error $\delta$ 
has query complexity $\Theta\rbra{\min\{n,\sqrt{n\log\rbra{1/\delta}}\}}$ \cite{BCWZ99}; 
and (iii) no dependence: there are promised exact cases in which the error parameter disappears:
in amplitude amplification, if the initial marked fraction is exactly $1/4$,
then a single Grover iterate succeeds with probability $1$ \cite{Gro96,BHMT02}.
This behavior is quite different from the classical case, and should be compared with the classical i.i.d. setting.
For testing two fixed distributions \(P\) and \(Q\), the optimal error exponent is governed by the Chernoff information \cite{Che52}; in the local regime, this leads to the familiar \(\Theta(\log(1/\delta)/d_{\mathrm H}^2(P,Q))\) scaling, as also reflected in the Hellinger distance lower bound framework of Bar-Yossef et al.\ \cite{BYKS01,BY02}.
At the same time, even in classical distribution testing, small error bounds are not always obtained by simply repeating a constant error tester.
For example, identity testing over an \(n\)-element domain has constant error sample complexity
\(\Theta(\sqrt n/\epsilon^2)\) \cite{Pan08,VV17}, whereas its optimal high-confidence complexity is
\(\Theta((\sqrt{n\log(1/\delta)}+\log(1/\delta))/\epsilon^2)\) \cite{DGPP18}.
The latter improves over black-box amplification of a constant error tester, which would require
\(O(\sqrt n\log(1/\delta)/\epsilon^2)\) samples \cite{DGPP18}.
More generally, recent work on high-probability distribution testing treats the dependence on \(\delta\) as part of the main complexity characterization \cite{DGKPP21}. 
Thus, in the quantum setting, where logarithmic, sub-logarithmic, and even absent \(\delta\)-dependence can all occur, it is natural to study small error quantum complexities with \(\delta\) treated as an intrinsic complexity parameter.

In this paper, we present a framework for quantum composite hypothesis testing with small error,
which allows us to systematically establish \textit{tight} quantum lower bounds
for a series of property testing and functional estimation tasks,
including $\ell_2$-closeness testing, Tsallis entropy estimation, and pure-state fidelity and trace distance estimation.

\subsection{Main results}
The central problem we study in this paper is the composite hypothesis testing for trigonometrically linear unitaries. 
A matrix $M\rbra{\theta}$ with parameter $\theta$ is said to be trigonometrically linear in $\theta$, if $M\rbra{\theta} = A e^{-i\theta} + B e^{i\theta} + C$ where $A$, $B$, and $C$ are matrices independent of $\theta$. 
This includes the oracle families used below, which consist of a phase gate composed with fixed unitaries.
Each query increases the trigonometric degree of the amplitudes by at most one. The Remez inequality bounds the degree needed for an acceptance probability to be at least $1-\delta$ on $\Theta_0$ and at most $\delta$ on $\Theta_1$.

\begin{definition}[Composite hypothesis testing for trigonometrically linear unitaries]
Let $\Theta := [-\pi,\pi)$, and let $\Theta_0,\Theta_1 \subseteq \Theta$ be two nonempty disjoint sets.
Given a quantum unitary oracle $U(\theta)$ that is trigonometrically linear in an unknown parameter $\theta \in \Theta$, the task $\textsc{HT}(\Theta_0,\Theta_1)$ is, under the promise that $\theta \in \Theta_0 \cup \Theta_1$, to determine which of the following two hypotheses holds:
\begin{itemize}
 \item Null hypothesis $H_0$:
    the parameter $\theta$ belongs to the null hypothesis set $\Theta_0$.
    \item Alternative hypothesis $H_1$:
    the parameter $\theta$ belongs to the alternative hypothesis set $\Theta_1$.
\end{itemize}
\end{definition}

Our main result is a lower bound for this composite hypothesis testing problem. 
\begin{theorem}[Quantum composite hypothesis lower bound] \label{thm:main}
    Let $\Theta_0, \Theta_1 \subseteq \Theta = [-\pi,\pi)$ be nonempty disjoint measurable hypothesis sets with $\abs{\Theta_1} \geq 3\pi/2$. 
    For any $\delta \in \rbra{0, 1/2}$, any quantum algorithm that solves $\textsc{HT}\rbra{\Theta_0, \Theta_1}$ with both Type-I and Type-II errors at most $\delta$ requires at least
    \[
    \frac{1}{8 \rbra*{\abs*{\Theta}-\abs*{\Theta_1}}} \log\rbra*{\frac{1-\delta}{\delta}}
    \]
    queries to $U\rbra{\theta}$,  where \(\abs{\cdot}\) denotes the Lebesgue measure on \([-\pi,\pi)\). 
\end{theorem}

\Cref{thm:main} is the main technical tool behind our applications, yielding a series of tight lower bounds for fundamental problems in quantum property testing and functional estimation; see the applications below and \Cref{sec:applications}.

\paragraph{Relation to Mande and de Wolf.}
Mande and de Wolf~\cite[Section~5]{MdW23} prove their small error phase estimation lower bound using trigonometric polynomials and the Remez inequality.
They consider the phase oracle $U_\theta=I-(1-e^{i\theta})\ketbra{0}{0}$ and distinguish $\theta=0$ from phases outside an interval around zero.
\Cref{thm:main} applies the same proof to trigonometrically linear unitary families and disjoint measurable hypothesis sets. The bound depends on the total measure of $S=\Theta\setminus\Theta_1$, which may be a union of intervals.
In particular, the $\ell_2$ and Tsallis examples below can also be treated by a symmetry reduction or a direct adaptation of their argument. Our theorem gives a common statement for these reductions and the other applications.

\subsection{Applications} 
We next present the main applications of our method; these are summarized in \Cref{tab:applications}.

\begin{table}[htb]
  \caption{Summary of quantum query complexities. The middle column gives constant error bounds from the cited algorithms; the Hellinger upper bound follows by applying square-root amplitude estimation~\cite{Wan24}. The last column gives the small error bounds proved here.}
  \label{tab:applications}
  \centering
  \footnotesize
  \renewcommand{\arraystretch}{1.22}
  \begin{tabularx}{\linewidth}{@{}
    >{\raggedright\arraybackslash}m{0.12\linewidth}
    >{\raggedright\arraybackslash}m{0.13\linewidth}
    >{\centering\arraybackslash}m{0.38\linewidth}
    >{\raggedleft\arraybackslash}m{0.20\linewidth}
    @{}}
    \toprule
    \multicolumn{2}{@{}l}{Problem}
    &
    \multicolumn{1}{c}{Prior results (constant error)}
    &
    \multicolumn{1}{c@{}}{This paper}
    \\
    \midrule

    \multicolumn{2}{@{}m{0.27\linewidth}}{\(\ell_2\)-closeness testing}
    &
    \(\Theta(1/\epsilon)\) \cite{LWL24}
    &
    \(\Theta\rbra{\log\rbra{1/\delta}/\epsilon}\)
    \\
    \midrule

    \multirow{2}{=}{Tsallis entropy estimation}
    &
    integer \(q\ge 2\)
    &
    \(\Theta(1/\epsilon)\) \cite{BCWdW01,EAO+02,BHMT02}
    &
    \(\Theta\rbra{\log\rbra{1/\delta}/\epsilon}\)
    \\
    \cmidrule(l){2-4}

    &
    real \(q\ge 1.5\)
    &
    \(\widetilde O(1/\epsilon)\) \cite{CGLWW26}
    &
    \(\widetilde\Theta\rbra{\log\rbra{1/\delta}/\epsilon}\)
    \\
    \midrule

    \multicolumn{2}{@{}m{0.27\linewidth}}{Pure-state Uhlmann fidelity and trace distance estimation}
    &
    \(\Theta(1/\epsilon)\) \cite{Wan24}
    &
    \(\Theta\rbra{\log\rbra{1/\delta}/\epsilon}\)
    \\
    \midrule
    \multicolumn{2}{@{}>{\raggedright\arraybackslash}m{0.27\linewidth}}{Identity testing in Hellinger distance (point-mass reference)}
    & \(O(1/\epsilon)\) via \cite{Wan24}
    & \(\Theta\rbra{\log(1/\delta)/\epsilon}\)
    \\
    \midrule
    \multicolumn{2}{@{}>{\raggedright\arraybackslash}m{0.27\linewidth}}{Fidelity to a pure state and \(\sqrt{\tr(\rho\sigma^2)}\) estimation}
    & \(\Theta(1/\epsilon)\) \cite{FW25}
    & \(\Theta\rbra{\log(1/\delta)/\epsilon}\)
    \\
    \bottomrule
  \end{tabularx}
\end{table}

\paragraph{$\ell_2$-closeness testing.}
The first application is $\ell_2$-closeness testing: given unknown distributions $p$ and $q$, the task is to distinguish, under a promise, whether $p=q$ or $\Abs{p-q}_2 \geq \epsilon$.
Here, for a distribution $p = \rbra{p_1, p_2, \dots, p_n}$, we assume query access to the quantum oracle $\mathcal{O}_p$ such that
\[
\mathcal{O}_p \ket{0} = \sum_{i=1}^n \sqrt{p_i} \ket{i} \ket{\varphi_i},
\]
where the $\ket{\varphi_i}$'s are normalized auxiliary states. For $\ell_2$-closeness testing, we are given such access to both $\mathcal{O}_p$ and $\mathcal{O}_q$. 
This model is the standard quantum input model for discrete distributions, which is known as the purified quantum query access model \cite{GL20}. 

\begin{corollary}
    The quantum query complexity of $\ell_2$-closeness testing with small error $\delta$ is $\Theta\rbra{\log\rbra{1/\delta}/\epsilon}$.
    In other words, there is a quantum algorithm for $\ell_2$-closeness testing with small error $\delta$ with query complexity $O\rbra{\log\rbra{1/\delta}/\epsilon}$ and any quantum algorithm for $\ell_2$-closeness testing with small error $\delta$ requires query complexity $\Omega\rbra{\log\rbra{1/\delta}/\epsilon}$. 
\end{corollary}
The constant error version of this problem was recently settled by Luo et al.\ \cite{LWL24}, who gave a succinct quantum tester with query complexity $O(1/\epsilon)$ and proved that its dependence on $\epsilon$ is optimal. 
Our result strengthens this optimality to the small error regime: it shows that the logarithmic dependence on $1/\delta$ is also necessary.
Thus the natural high-confidence version of the tester of \cite{LWL24} is optimal in its joint dependence on $\epsilon$ and $\delta$.

\paragraph{Tsallis entropy estimation.}
The second application is $q$-Tsallis entropy estimation.
Under the same purified quantum query-access model, given query access to $\mathcal{O}_p$ for an unknown distribution $p$ over $[n]$, the task is to estimate the power sum
$F_q\rbra{p} = \sum_{i=1}^n p_i^q$
to within additive error $\epsilon$.
Estimating $F_q(p)$ is equivalent, up to an affine transformation, to estimating the $q$-Tsallis entropy
$\mathrm{S}_q(p) = \frac{1-F_q(p)}{q-1}$;
in particular, the case $q=2$ corresponds to the Gini impurity.
We obtain the following small error query complexity for this problem.
\begin{corollary}
    The quantum query complexity of $q$-Tsallis entropy estimation with small error $\delta$ is $\widetilde{\Theta}\rbra{\log\rbra{1/\delta}/\epsilon}$ for any $q \geq 1.5$.\footnote{$\widetilde{\Theta}\rbra{\cdot}$ suppresses polylogarithmic factors in $\epsilon$.}
    Moreover, if $q \geq 2$ is integer, then the complexity is $\Theta\rbra{\log\rbra{1/\delta}/\epsilon}$.
\end{corollary}
For integer $q \geq 2$, the functional $F_q(p)$ is a power sum and can be estimated using swap-test and cyclic-permutation primitives, as in the approaches of Buhrman et al.\ \cite{BCWdW01} and Ekert et al.\ \cite{EAO+02}; combined with amplitude estimation~\cite{BHMT02}, these give the matching upper bound.
Our corollary proves that these integer-order algorithms are optimal, including their dependence on the small error probability $\delta$.
For real orders $q \geq 1.5$, the recent quantum multi-level estimation framework of Chen et al.\ \cite{CGLWW26} gives near-optimal algorithms for such distribution functionals.
Our lower bound shows that, in this range of $q$, their dependence on $\epsilon$ and $\delta$ is optimal up to polylogarithmic factors.

\paragraph{Pure-state fidelity and trace distance estimation.}
Our framework also applies to estimation problems for quantum objects.
The third application concerns two pure-state closeness measures: Uhlmann fidelity and trace distance.
Given state-preparation query access to two unknown pure states $\ket{\varphi}$ and $\ket{\phi}$, namely quantum unitary oracles $\mathcal{O}_{\varphi}$ and $\mathcal{O}_{\phi}$ such that
$\mathcal{O}_{\varphi}\ket{0}=\ket{\varphi}$ and $\mathcal{O}_{\phi}\ket{0}=\ket{\phi}$,
the tasks are to estimate, to additive error $\epsilon$, the Uhlmann fidelity 
$\mathrm{F}\rbra{\ket{\phi},\ket{\varphi}} = \abs{\braket{\varphi}{\phi}}$
and the trace distance
$\mathrm{T}\rbra{\ket{\phi},\ket{\varphi}} = \sqrt{1-\abs{\braket{\varphi}{\phi}}^2}$.
We obtain the following tight small error complexity for both estimation tasks.

\begin{corollary}
    The quantum query complexity of both pure-state Uhlmann fidelity estimation and trace distance estimation with small error $\delta$ is $\Theta\rbra{\log\rbra{1/\delta}/\epsilon}$. 
\end{corollary}

For constant error, Wang~\cite{Wan24} developed separate optimal quantum algorithms for pure-state fidelity estimation and trace distance estimation, each using $\Theta(1/\epsilon)$ queries to the state-preparation circuits and quadratically improving the folklore $O(1/\epsilon^2)$ approaches based on SWAP tests.
Our corollary extends these constant error optimality results to the small error regime, showing that each of the two estimation tasks has query complexity $\Theta\rbra{\log\rbra{1/\delta}/\epsilon}$.

\paragraph{Further applications.}
We also prove a tight small error bound for identity testing in Hellinger distance when the reference distribution is a point mass, with an equivalent statement in terms of the Bhattacharyya coefficient (\Cref{sec:hellinger}).
Our pure-state fidelity lower bound also gives matching small error bounds for estimating fidelity to a pure state and $\sqrt{\tr(\rho\sigma^2)}$, extending the constant error results of Fang and Wang~\cite{FW25} (\Cref{sec:mixed-fidelity}).

\subsection{Techniques}

\paragraph{Composite testing through trigonometric query families.}
Our lower bound framework is inspired by the classical theory of composite hypothesis testing, which has long served as a systematic source of information-theoretic lower bounds in statistics. 
We bring this perspective into the quantum query setting by formulating a composite testing problem for trigonometrically linear unitary families. 
Concretely, rather than reducing a problem to two isolated hard instances, we embed it into a one-parameter family of quantum oracles $\{U(\theta)\}_{\theta\in\Theta}$ and ask an algorithm to distinguish a null parameter set $\Theta_0$ from a large alternative parameter set $\Theta_1$. 
This formulation is particularly well suited to the small error regime: the lower bound is governed not merely by a pairwise separation between hard instances, but by how sharply a quantum algorithm can separate $\Theta_0$ from $\Theta_1$ when $\Theta_1$ occupies almost all of the parameter space.
Classical lower bounds based on Chernoff information concern i.i.d. observations. In the quantum query model, an algorithm may interleave coherent oracle calls with arbitrary unitaries; we control such algorithms by bounding the degree of their acceptance probability.

The reason this composite viewpoint becomes powerful in the quantum setting is that many standard quantum oracles depend on an underlying phase parameter through trigonometric polynomials.
For the oracle families used in our applications, every matrix entry of $U(\theta)$ is trigonometrically linear in $\theta$. 
It follows from the quantum polynomial method \cite{BBC+01} that the acceptance probability of any $T$-query quantum algorithm is a real trigonometric polynomial of degree at most $2T$ (cf.\ \cite{MdW23}). 
Thus, a successful small error tester would give a low-degree trigonometric polynomial that is close to one on the null side and close to zero on almost all of the parameter space. 
The exponential-form Remez inequality (see \Cref{thm:trig-magic}) shows that such a separation is impossible unless the degree grows logarithmically in $1/\delta$ and inversely with the measure of the parameter region not covered by the alternative hypothesis. 
In the notation of our main theorem, this yields a lower bound of order $\Omega(\log(1/\delta)/(|\Theta|-|\Theta_1|))$.

\paragraph{Instantiating the framework.}
To apply the framework, the main task is to embed a concrete testing or estimation problem into a one-parameter trigonometrically linear oracle family and to choose the hypothesis sets so that the alternative region occupies almost all of \(\Theta\). 
Let us illustrate this with \(\ell_2\)-closeness testing. 
We fix \(p^\star=(1/2,1/2,0,\dots,0)\), and let the second distribution vary as \(q^{(\theta)}=(\cos^2(\theta/2),\sin^2(\theta/2),0,\dots,0)\), which is generated by the Hadamard--phase--Hadamard unitary \(W_\theta=H\operatorname{diag}(1,e^{i\theta})H\). 
This gives a trigonometrically linear oracle family, and the \(\ell_2\) distance satisfies \(\|q^{(\theta)}-p^\star\|_2=|\cos(\theta)|/\sqrt2\). 
Thus the null case occurs at \(\theta^\star=\pi/2\). 
We choose \(\Theta_0=\{\pi/2\}\) and \(\Theta_1=\{\theta:\|q^{(\theta)}-p^\star\|_2\ge\epsilon\}\). 
The exceptional region outside \(\Theta_1\) is exactly the set where \(|\cos(\theta)|<\sqrt2\,\epsilon\), whose measure is \(\Theta(\epsilon)\). 
Therefore, \Cref{thm:main} gives the lower bound \(\Omega(\log(1/\delta)/\epsilon)\). 
The other applications follow the same pattern: for Tsallis entropy estimation, we choose a two-point family whose power sum varies linearly near a reference phase; for pure-state fidelity and trace distance estimation, the Hadamard--phase--Hadamard family makes the relevant quantities simple trigonometric functions of \(\theta\). 
In each case, the exceptional region has measure \(O(\epsilon)\), and the composite-testing lower bound yields the desired small error dependence.

\subsection{Discussion}

We developed a quantum composite hypothesis testing framework for trigonometrically linear unitary families, turning small exceptional sets in phase space into tight small error quantum lower bounds. 
Together with known algorithms, this framework settles the small error complexities of several quantum testing and estimation problems considered in this paper.

Several directions remain open. 
First, it would be interesting to develop a quantum analogue of the method of two fuzzy hypotheses \cite{Tsy09}. 
Our present framework separates a null parameter set from a large alternative set along a deterministic one-parameter oracle family. 
In contrast, classical minimax lower-bound theory contains several complementary tools, including Le Cam's method, Assouad's lemma, Fano's method, and the method of two fuzzy hypotheses, which use carefully chosen finite sets, hypercubes, packings, or priors over composite hypotheses to prove lower bounds \cite{LC73,Yu97,Tsy09,JHW15}. 
A quantum fuzzy-hypothesis principle would aim to compare the behavior of quantum algorithms under two distributions over oracle instances or states, rather than under a single localized exceptional set. 
Such a principle could handle hard instances that arise from averaging over structured families of oracles or states, and may bring quantum minimax lower bounds closer to their classical counterparts.

Second, our framework is intrinsically one-dimensional: the hard family is parametrized by a single phase \(\theta\), and the lower bound is controlled by the measure of an exceptional subset of \([-\pi,\pi)\). 
Many high-dimensional statistical problems have hard instances with several independent perturbation directions. 
A natural next step is to develop multi-parameter extensions of the present approach. 
For phase-parametrized oracle families, this would lead to multivariate trigonometric polynomials; for other parametrizations, ordinary or structured multivariate polynomials may be more appropriate. 
The main challenge is to find higher-dimensional analogues of the Remez principle that convert the geometry of exceptional sets into quantum query lower bounds.

Third, our results reinforce the view that the dependence on the error probability \(\delta\) should be treated as an intrinsic part of the complexity, rather than merely as a black-box amplification factor. 
This phenomenon is already familiar in classical distribution testing: the optimal high-confidence complexity of identity testing improves over naive amplification of the constant error tester \cite{DGPP18}. 
In the quantum setting, the landscape appears even richer. 
The problems studied in this paper have logarithmic dependence on \(1/\delta\), as in quantum phase estimation \cite{MdW23}, while quantum search exhibits a different high-confidence dependence on \(\delta\) \cite{BCWZ99}. 
For a fixed oracle family and exceptional set $S$, \Cref{thm:main} gives a lower bound proportional to $\log((1-\delta)/\delta)/|S|$.
This dependence does not explain the $\sqrt{\log(1/\delta)}$ factor in the optimal complexity of quantum search~\cite{BCWZ99}.

Understanding which quantum testing and estimation problems admit small error behavior beyond black-box amplification remains an interesting direction.

Finally, the present applications all rely on relatively simple phase-parametrized hard families. 
It would be useful to identify other settings in which such families exist and in which the relevant quantity has controlled local variation near a reference phase. 
This may lead to further small error lower bounds for related quantum statistical estimation problems.

\section{Preliminaries}

We briefly recall the basic quantum notation and the query model used throughout the paper. 
A qubit is a unit vector in \(\mathbb{C}^2\), and an \(n\)-qubit pure state is a unit vector in \((\mathbb{C}^2)^{\otimes n}\). 
We use Dirac notation and write the computational basis as \(\{\ket{x}:x\in\{0,1\}^n\}\). 
If a state \(\ket{\psi}=\sum_x \alpha_x\ket{x}\) is measured in the computational basis, the outcome is \(x\) with probability \(|\alpha_x|^2=|\bra{x}\psi\rangle|^2\).

A quantum oracle is a unitary operator \(\mathcal O\) on a finite-dimensional Hilbert space. 
A \(T\)-query quantum algorithm \(\mathcal A\) with oracle access to \(\mathcal O\) consists of oracle-independent unitaries \(U_0,U_1,\dots,U_T\), together with \(T\) oracle calls \(\mathcal O_1,\dots,\mathcal O_T\), where each \(\mathcal O_j\) may be \(\mathcal O\), \(\mathcal O^\dagger\), or a controlled version thereof. 
Starting from a fixed initial state \(\ket{0}\), the algorithm produces the final state \(\ket{\psi}=U_T\mathcal O_TU_{T-1}\cdots U_1\mathcal O_1U_0\ket{0}\). 
It then performs a final measurement in the computational basis and outputs \(x\) with probability \(\Pr[\mathcal A^{\mathcal O}=x]=|\bra{x}\psi\rangle|^2\). 
More generally, for an event \(\Gamma\) consisting of several possible outcomes, we write \(\Pr[\mathcal A^{\mathcal O}\in\Gamma]=\sum_{x\in\Gamma}|\bra{x}\psi\rangle|^2\). 
The query complexity of \(\mathcal A\) is the number \(T\) of oracle calls.

\section{Quantum Composite Hypothesis Testing}
\label{sec:main-thm}
In this section, we prove the main composite-testing lower bound, \Cref{thm:main}. 
Let \(\Theta=[-\pi,\pi)\). 
For a measurable set \(E\subseteq\Theta\), we write \(\abs*{E}\) for its Lebesgue measure. 
We say that a function \(p:\Theta\to\mathbb C\) is a trigonometric polynomial of degree at most \(d\) if it can be written as \(p(\theta)=\sum_{k=-d}^{d} c_k e^{ik\theta}\).

The proof can be viewed as a trigonometric analogue of the quantum polynomial method. 
Along a trigonometrically linear unitary family \(U(\theta)\), any \(T\)-query algorithm has acceptance probability given by a real trigonometric polynomial of degree at most \(2T\). 
Thus, a small error algorithm for the composite testing problem would induce a low-degree trigonometric polynomial separating the null hypothesis set from a large alternative hypothesis set. 
The following Remez-type inequality quantifies the obstruction to such a separation.
\begin{theorem}[Exponential-form Remez inequality, {\cite[Theorem~5.1.2]{BE95}}]
\label{thm:trig-magic}
Let \(p:\Theta\to\mathbb R\) be a real trigonometric polynomial of degree at most \(d\), and let \(E\subseteq\Theta\) be measurable with \(\abs*{E}\le \pi/2\). 
Then
\[
    \sup_{\theta\in\Theta}\abs*{p(\theta)}
    \le
    \exp\rbra*{4d\abs*{E}}
    \sup_{\theta\in\Theta\setminus E}\abs*{p(\theta)} .
\]
\end{theorem}
We now turn to the proof of \Cref{thm:main}.
\begin{proof}[Proof of \Cref{thm:main}]
Let \(\mathcal A\) be a \(T\)-query quantum algorithm that solves \(\textsc{HT}(\Theta_0,\Theta_1)\) with both Type-I and Type-II errors at most \(\delta\). 
Let \(\Gamma\) be the event that \(\mathcal A\) outputs the null hypothesis \(H_0\). 
For each \(\theta\in\Theta\), define \(f(\theta)\coloneqq \Pr[\mathcal A^{U(\theta)}\in\Gamma]\).

We first record the trigonometric-polynomial structure of \(f\), following the degree-counting argument of~\cite[Section~5]{MdW23}. 
By deferred measurement, we may assume that \(\mathcal A\) performs only a final measurement. 
After \(t\) oracle calls, every basis-state amplitude is a trigonometric polynomial of degree at most \(t\) in \(\theta\). 
Indeed, this is trivial for \(t=0\), and each query to \(U(\theta)\), \(U(\theta)^\dagger\), or a controlled version thereof increases the degree by at most one, since the entries of \(U(\theta)\) are trigonometrically linear in \(\theta\); the oracle-independent unitaries do not change the degree. 
Thus every final amplitude \(\alpha_x(\theta)\) has degree at most \(T\), and hence
\[
    f(\theta)=\sum_{x\in\Gamma}\abs*{\alpha_x(\theta)}^2
\]
is a real trigonometric polynomial of degree at most \(2T\). Indeed, multiplying frequencies $e^{ik\theta}$ and $e^{-i\ell\theta}$ from an amplitude and its conjugate produces only frequencies $k-\ell\in[-2T,2T]$.

Now choose any reference point \(\theta^\star\in\Theta_0\), and set \(S=\Theta\setminus\Theta_1\). 
Since \(\Theta_0\cap\Theta_1=\emptyset\), we have \(\theta^\star\in S\). 
Moreover, the assumption \(\abs*{\Theta_1}\ge 3\pi/2\) gives \(\abs*{S}=\abs*{\Theta}-\abs*{\Theta_1}\le \pi/2\).
The Type-I error guarantee gives \(f(\theta^\star)\ge 1-\delta\), while the Type-II error guarantee gives \(f(\theta)\le \delta\) for every \(\theta\in\Theta_1=\Theta\setminus S\).

Apply \Cref{thm:trig-magic} to the real trigonometric polynomial \(g(\theta)=f(\theta)/\delta\), whose degree is at most \(2T\). 
Since \(\abs*{g(\theta)}\le 1\) for every \(\theta\in\Theta\setminus S\), we obtain
\[
    \frac{1-\delta}{\delta}
    \le
    g(\theta^\star)
    \le
    \sup_{\theta\in\Theta}\abs*{g(\theta)}
    \le
    \exp\rbra*{8T\abs*{S}} .
\]
Rearranging yields
\[
    T
    \ge
    \frac{1}{8\abs*{S}}
    \log\rbra*{\frac{1-\delta}{\delta}}
    =
    \frac{1}{8\rbra*{\abs*{\Theta}-\abs*{\Theta_1}}}
    \log\rbra*{\frac{1-\delta}{\delta}},
\]
as claimed.
\end{proof}

\section{Applications}
\label{sec:applications}

We now instantiate \Cref{thm:main} for several testing and estimation problems. 
In each application, the key step is a carefully chosen one-parameter trigonometrically linear oracle family that converts the desired testing or estimation task into a composite hypothesis testing problem. 
We then choose hypothesis sets \(\Theta_0,\Theta_1\subseteq\Theta\) so that the alternative set covers all but a small exceptional region, and \Cref{thm:main} yields the desired lower bound.
\subsection{\texorpdfstring{\(\ell_2\)-closeness testing}{l2-closeness testing}}
We first consider \(\ell_2\)-closeness testing. 
Our hard family fixes one distribution and varies the other along a two-dimensional phase family for which the \(\ell_2\) distance is linear around the null phase.

\begin{theorem}[Quantum lower bound for \(\ell_2\)-closeness testing]
\label{thm:l2-closeness-lb}
For every \(\epsilon\in(0,1/4]\), \(\delta\in(0,1/3]\), and \(n\ge 2\), consider the following promise problem: given purified quantum query access to two unknown distributions \(p,q\) on \([n]\), decide whether
\[
p = q \quad \textup{or} \quad \Abs{p-q}_2 \geq \varepsilon
\]
with success probability at least \(1-\delta\). 
Then every quantum algorithm for this problem requires \(\Omega\rbra*{\log(1/\delta)/\epsilon}\) queries.
\end{theorem}

\begin{proof}
Fix \(p^\star=(1/2,1/2,0,\dots,0)\). 
Let 
\[
H = \frac{1}{\sqrt{2}} \begin{pmatrix}
    1 & 1 \\
    1 & -1
\end{pmatrix}
\]
be the Hadamard gate, and set \(W_\theta=H\operatorname{diag}(1,e^{i\theta})H\) and \(\widetilde W_\theta=W_\theta\oplus I_{n-2}\).

Let \(q^{(\theta)}\) be the distribution prepared by the purified oracle \(\widetilde W_\theta\). 
Then \(q^{(\theta)}=(\cos^2(\theta/2),\sin^2(\theta/2),0,\dots,0)\), and every entry of \(\widetilde W_\theta\) is trigonometrically linear in \(\theta\).

Define \(g(\theta)=\Abs*{q^{(\theta)}-p^\star}_2\). 
A direct calculation gives \(g(\theta)=|\cos\theta|/\sqrt 2\). 
We choose \(\Theta_0=\{\pi/2\}\) and \(\Theta_1=\{\theta\in\Theta:g(\theta)\ge \epsilon\}\). 
Then \(\Theta_0\cap\Theta_1=\emptyset\), and the exceptional region is \(\Theta\setminus\Theta_1=\{\theta\in\Theta:|\cos\theta|<\sqrt2\,\epsilon\}\).
Hence \(\abs*{\Theta}-\abs*{\Theta_1}=4\arcsin(\sqrt2\,\epsilon)=O(\epsilon)\). 
For \(\epsilon\le 1/4\), this measure is at most \(4\arcsin(\sqrt2/4)<\pi/2\), so \(\abs*{\Theta_1}\ge 3\pi/2\).

Now suppose there is a \(T\)-query tester for the original \(\ell_2\)-closeness problem. 
Restricting it to the family \((p^\star,q^{(\theta)})\) gives an algorithm for \(\textsc{HT}(\Theta_0,\Theta_1)\): it outputs \(H_0\) when the tester declares \(p=q\), and \(H_1\) otherwise. 
The oracle for \(p^\star\) is fixed and can be absorbed into the oracle-independent part of the circuit, so the only parameter-dependent oracle is \(\widetilde W_\theta\). 
By \Cref{thm:main}, 
\[
T\ge \Omega\rbra*{\frac{1}{\abs*{\Theta}-\abs*{\Theta_1}}\log\rbra*{\frac{1}{\delta}}}=\Omega\rbra*{\frac{1}{\varepsilon}\log\rbra*{\frac{1}{\delta}}}.
\]
\end{proof}
The distance vanishes at $\theta=\pm\pi/2$ and grows linearly near both points, so the excluded set consists of two intervals of total length $O(\epsilon)$.

\subsection{Tsallis entropy estimation}
We next consider \(q\)-Tsallis entropy estimation. 
For fixed \(q>0\) with \(q\neq1\), estimating the \(q\)-Tsallis entropy \(S_q(p)=(1-F_q(p))/(q-1)\) is equivalent, up to a known affine transformation, to estimating the power sum \(F_q(p)=\sum_i p_i^q\).
The case \(q=2\), corresponding to the Gini impurity, already illustrates the construction: on a two-point family \(p^{(\theta)}=(x(\theta),1-x(\theta))\) with \(x(\theta)=u_0+v_0\sin\theta\) and \(u_0\neq 1/2\), the quantity \(F_2(p^{(\theta)})=x(\theta)^2+(1-x(\theta))^2\) has a nonzero derivative at the reference phase.
The same idea works for every fixed \(q>0\), \(q\neq 1\), by keeping \(x(\theta)\) inside an interval contained in \((1/2,1)\), where the derivative of \(x^q+(1-x)^q\) is bounded away from zero.

\begin{theorem}[Quantum lower bound for estimating \(F_q\)]
\label{thm:Tsallis-small-error-lb}
Fix \(q> 0\) with \(q\neq 1\). 
There exists a constant \(c_q>0\) such that the following holds. 
For every \(\epsilon\in(0,c_q]\), \(\delta\in(0,1/3]\), and \(n\ge 2\), consider the problem of estimating \(F_q(p)=\sum_{i=1}^n p_i^q\) to additive error \(\epsilon\) with success probability at least \(1-\delta\), given purified quantum query access to an unknown distribution \(p\) on \([n]\). 
Then every quantum algorithm for this problem requires \(\Omega\rbra*{\log(1/\delta)/\epsilon}\) queries.
\end{theorem}
\begin{proof}
We construct a two-point family \(p^{(\theta)}=(x(\theta),1-x(\theta),0,\dots,0)\) whose first coordinate stays inside a fixed interval contained in \((1/2,1)\). 
Let \(\phi=\pi/10\), \(\psi=\pi/20\), \(D_\theta=\operatorname{diag}(1,e^{i\theta})\), and
\[
    A=
    \begin{pmatrix}
    \cos\phi & -i\sin\phi\\
    -i\sin\phi & \cos\phi
    \end{pmatrix},
    \qquad
    B=
    \begin{pmatrix}
    \cos\psi & -\sin\psi\\
    \sin\psi & \cos\psi
    \end{pmatrix}.
\]
Set \(W_\theta=AD_\theta B\) and \(\widetilde W_\theta=W_\theta\oplus I_{n-2}\), and let \(p^{(\theta)}\) be the distribution prepared by \(\widetilde W_\theta\). 
The family \(\widetilde W_\theta\) is trigonometrically linear. 
Writing \(x(\theta)=p_1^{(\theta)}\), a direct calculation gives \(x(\theta)=u_0+v_0\sin\theta\), where
\begin{align*}
    u_0&=\cos^2\phi\cos^2\psi+\sin^2\phi\sin^2\psi, \\
    v_0&=2\cos\phi\cos\psi\sin\phi\sin\psi>0.
\end{align*}
Moreover,
\[
    x(\theta)\in I
    \coloneqq
    [\cos^2(3\pi/20),\cos^2(\pi/20)]
    \subset(1/2,1)
\]
for every \(\theta\in\Theta\).

Let \(h_q(x)=x^q+(1-x)^q\); then \(F_q(p^{(\theta)})=h_q(x(\theta))\). 
For \(x\in I\), we have \(x>1-x\). 
Consider the following two cases.
\begin{itemize}
    \item If \(q>1\), then \(t\mapsto t^{q-1}\) is strictly increasing, so
\(h_q'(x)=q(x^{q-1}-(1-x)^{q-1})>0\) on \(I\). 
    \item If \(0<q<1\), then \(t\mapsto t^{q-1}\) is strictly decreasing, so \(h_q'(x)<0\) on \(I\). 
\end{itemize}
In either case, \(h_q'\) has no zero on \(I\). 
Since \(h_q'\) is continuous on the compact interval \(I\), we have \(m_q\coloneqq\min_{x\in I}|h_q'(x)|>0\).
Set \(\kappa_q=m_qv_0\). 
By the mean value theorem,
\[
    |F_q(p^{(\theta)})-F_q(p^{(0)})|
    =
    |h_q(x(\theta))-h_q(x(0))|
    \ge
    m_q|x(\theta)-x(0)|
    =
    \kappa_q|\sin\theta|.
\]

Set \(c_q=\kappa_q/8\), and assume \(\epsilon\le c_q\). 
Define
\[
    S=\{\theta\in\Theta:|F_q(p^{(\theta)})-F_q(p^{(0)})|\le 2\epsilon\}.
\]
The previous inequality implies \(S\subseteq\{\theta:|\sin\theta|\le 2\epsilon/\kappa_q\}\), and hence \(\abs*{S}\le 4\arcsin(2\epsilon/\kappa_q)=O_q(\epsilon)\). 
Also, since \(2\epsilon/\kappa_q\le 1/4\), we have \(\abs*{S}\le 4\arcsin(1/4)<\pi/2\).

Take \(\Theta_0=\{0\}\) and \(\Theta_1=\Theta\setminus S\). 
Suppose a \(T\)-query estimator \(\widetilde F\) is \(\epsilon\)-accurate for \(F_q\) with success probability at least \(1-\delta\). 
We turn it into a tester for \(\textsc{HT}(\Theta_0,\Theta_1)\) by outputting \(H_0\) if and only if \(|\widetilde F-F_q(p^{(0)})|\le \epsilon\). 
At \(\theta=0\), this event occurs with probability at least \(1-\delta\). 
For every \(\theta\in\Theta_1\), we have \(|F_q(p^{(\theta)})-F_q(p^{(0)})|>2\epsilon\), so by the triangle inequality an \(\epsilon\)-accurate estimate cannot lead to output \(H_0\). 
Thus the Type-II error is at most \(\delta\). 
By \Cref{thm:main}, 
\[
T\ge \Omega_q\rbra*{\frac{1}{\abs*{S}}\log\rbra*{\frac{1}{\delta}}}=\Omega_q\rbra*{\frac{1}{\varepsilon}\log\rbra*{\frac{1}{\delta}}}.
\]
\end{proof}
The choice $I\subset(1/2,1)$ ensures that $|h_q'|$ is bounded below by a positive constant. Hence the phases for which $|F_q(p^{(\theta)})-F_q(p^{(0)})|\le2\epsilon$ lie in intervals of total length $O_q(\epsilon)$ around $0$ and $\pi$ modulo $2\pi$.

\subsection{Pure-state fidelity and trace distance estimation}
We next apply the framework to pure-state functional estimation. 
A single two-dimensional phase family suffices for both Uhlmann fidelity \(\mathrm{F}(\ket{\phi},\ket{\varphi})=\abs{\braket{\varphi}{\phi}}\) and trace distance \(\mathrm{T}(\ket{\phi},\ket{\varphi})=\sqrt{1-\abs{\braket{\varphi}{\phi}}^2}\), since both quantities are simple functions of the overlap between the two states.

\begin{theorem}[Quantum lower bound for pure-state Uhlmann fidelity estimation]
\label{thm:fidelity-small-error-lb}
For every \(\epsilon\in(0,1/32]\) and \(\delta\in(0,1/3]\), consider the following problem: given state-preparation query access to two unknown pure states \(\ket{\phi}\) and \(\ket{\varphi}\), output an estimate \(\widetilde F\) satisfying \(|\widetilde F-\mathrm{F}\rbra{\ket{\phi},\ket{\varphi}}|\le \epsilon\) with success probability at least \(1-\delta\). 
Then every quantum algorithm for this problem requires \(\Omega\rbra*{\log(1/\delta)/\epsilon}\) queries.
\end{theorem}

\begin{proof}
Fix \(\ket{\phi}=\ket{0}\), let \(W_\theta=H\operatorname{diag}(1,e^{i\theta})H\), and set \(\ket{\varphi^{(\theta)}}=W_\theta\ket{0}\). 
The oracle preparing \(\ket{\phi}\) is fixed, while \(W_\theta\) is trigonometrically linear. 
For \(\theta\in[-\pi,\pi)\), the Uhlmann fidelity along this family is
\[
    F(\theta)
    =
    \left |\braket{\phi}{\varphi^{(\theta)}}\right|
    =
    |\bra{0}W_\theta\ket{0}|
    =
    \cos(\theta/2).
\]

Take 
\[
\theta^\star=\pi/2, \quad a=F(\theta^\star)=1/\sqrt2, \quad S=\{\theta\in\Theta:|F(\theta)-a|\le 2\epsilon\}.
\]
For \(u=2\epsilon\le 1/16\), the set \(S\) has measure \(4(\arccos(a-u)-\arccos(a+u))\). 
Since the derivative of \(\arccos\) is bounded on \([a-1/16,a+1/16]\), we get \(\abs*{S}=O(\epsilon)\); moreover, for \(\epsilon\le 1/32\), this measure is at most \(\pi/2\). 
Let \(\Theta_0=\{\theta^\star\}\) and \(\Theta_1=\Theta\setminus S\).

Suppose a \(T\)-query estimator \(\widetilde F\) is \(\epsilon\)-accurate with probability at least \(1-\delta\). 
Define a tester by outputting \(H_0\) if and only if \(|\widetilde F-a|\le \epsilon\). 
At \(\theta=\theta^\star\), this happens with probability at least \(1-\delta\). 
For every \(\theta\in\Theta_1\), we have \(|F(\theta)-a|>2\epsilon\), so an \(\epsilon\)-accurate estimate cannot lead to output \(H_0\). 
Therefore the tester solves \(\textsc{HT}(\Theta_0,\Theta_1)\) with error at most \(\delta\). 
By \Cref{thm:main}, 
\[
T\ge \Omega\rbra*{\frac{1}{\abs*{S}}\log\rbra*{\frac{1}{\delta}}}=\Omega\rbra*{\frac{1}{\varepsilon}\log\rbra*{\frac{1}{\delta}}}.
\]
\end{proof}
We choose the reference value $1/\sqrt2$ because $F'(\theta)\ne0$ at both phases $\theta=\pm\pi/2$ where it is attained. This gives $|S|=O(\epsilon)$.

\begin{theorem}[Quantum lower bound for pure-state trace distance estimation]
\label{thm:trace-small-error-lb}
For every \(\epsilon\in(0,1/8]\) and  \(\delta\in(0,1/3]\), consider the following problem: given state-preparation query access to two unknown pure states \(\ket{\phi}\) and \(\ket{\varphi}\), output an estimate \(\widetilde T\) satisfying \(|\widetilde T-\mathrm{T}\rbra{\ket{\phi},\ket{\varphi}}|\le \epsilon\) with success probability at least \(1-\delta\).
Then every quantum algorithm for this problem requires \(\Omega\rbra*{\log(1/\delta)/\epsilon}\) queries.
\end{theorem}

\begin{proof}
Use the same family as above: \(\ket{\phi}=\ket{0}\) and \(\ket{\varphi^{(\theta)}}=W_\theta\ket{0}\), where \(W_\theta=H\operatorname{diag}(1,e^{i\theta})H\).
The trace distance along this family is \(D(\theta)=\mathrm{T}(\ket{\phi},\ket{\varphi^{(\theta)}})=|\sin(\theta/2)|\).
Take 
\[
\theta^\star=0, \quad S=\{\theta\in\Theta:D(\theta)\le 2\epsilon\}, \quad \Theta_0=\{0\}, \quad \Theta_1=\Theta\setminus S.
\]
For \(\epsilon\le 1/8\), we have
\[
    \abs*{S}
    =
    \abs*{\{\theta\in\Theta:|\sin(\theta/2)|\le 2\epsilon\}}
    =
    4\arcsin(2\epsilon)
    =
    O(\epsilon),
\]
and \(\abs*{S}\le 4\arcsin(1/4)<\pi/2\).

Given an \(\epsilon\)-accurate trace distance estimator \(\widetilde T\), define a tester that outputs \(H_0\) if and only if \(\widetilde T\le \epsilon\). 
At \(\theta=0\), the true trace distance is \(0\), so the tester outputs \(H_0\) with probability at least \(1-\delta\). 
For every \(\theta\in\Theta_1\), the true trace distance is larger than \(2\epsilon\), so an \(\epsilon\)-accurate estimate cannot lead to output \(H_0\). 
Therefore the tester solves \(\textsc{HT}(\Theta_0,\Theta_1)\) with error at most \(\delta\). 
By \Cref{thm:main}, 
\[
T\ge \Omega\rbra*{\frac{1}{\abs*{S}}\log\rbra*{\frac{1}{\delta}}}=\Omega\rbra*{\frac{1}{\varepsilon}\log\rbra*{\frac{1}{\delta}}}.
\]
\end{proof}

Here $D(\theta)=|\sin(\theta/2)|$ grows linearly in $|\theta|$ near zero, which gives $|S|=O(\epsilon)$.

\subsection{Identity testing in Hellinger distance}
\label{sec:hellinger}
For distributions $p,r$, let $\mathrm{BC}(p,r)=\sum_i\sqrt{p_i r_i}$ be the Bhattacharyya coefficient.
We use the normalized Hellinger distance $d_{\mathrm H}(p,r)=\bigl(\frac12\sum_i(\sqrt{p_i}-\sqrt{r_i})^2\bigr)^{1/2}=\sqrt{1-\mathrm{BC}(p,r)}$.
We consider identity testing with the known reference distribution $p^\star=(1,0,\ldots,0)$ and purified quantum query access to $r$.

\begin{corollary}
\label{cor:hellinger}
For $n\ge2$, $0<\delta\le1/3$, and $0<\epsilon\le1/8$, distinguishing $r=p^\star$ from $d_{\mathrm H}(r,p^\star)\ge\epsilon$ with error probability at most $\delta$ has query complexity $\Theta(\log(1/\delta)/\epsilon)$.
Equivalently, for $0<\eta\le1/64$, distinguishing $\mathrm{BC}(r,p^\star)=1$ from $\mathrm{BC}(r,p^\star)\le1-\eta$ has complexity $\Theta(\log(1/\delta)/\sqrt\eta)$.
\end{corollary}
\begin{proof}
For the lower bound, use $W_\theta=H\operatorname{diag}(1,e^{i\theta})H$, embedded in $n$ dimensions as in the proof of \Cref{thm:l2-closeness-lb}, to prepare $r^{(\theta)}=(\cos^2(\theta/2),\sin^2(\theta/2),0,\ldots,0)$.
On $\Theta=[-\pi,\pi)$,
\[
 \mathrm{BC}(r^{(\theta)},p^\star)=\cos(\theta/2),
 \qquad d_{\mathrm H}(r^{(\theta)},p^\star)=\sqrt2\,|\sin(\theta/4)|.
\]
Take $\Theta_0=\{0\}$ and $\Theta_1=\{\theta:d_{\mathrm H}(r^{(\theta)},p^\star)\ge\epsilon\}$.
Then $|\Theta\setminus\Theta_1|=8\arcsin(\epsilon/\sqrt2)=O(\epsilon)$, and this measure is at most $\pi/2$ for $\epsilon\le1/8$.
The lower bound follows from \Cref{thm:main}.

For the upper bound, let $h=d_{\mathrm H}(r,p^\star)$ and let $a=1-r_1$ be the probability of observing an outcome other than the first.
Since $a=h^2(2-h^2)$, the promise implies either $a=0$ or $\sqrt a\ge\epsilon$.
Square-root amplitude estimation~\cite{Wan24} estimates $\sqrt a$ to accuracy $\epsilon/3$ using $O(1/\epsilon)$ queries at constant error. Comparing the estimate with $\epsilon/2$ decides the two cases; $O(\log(1/\delta))$ independent repetitions and a majority vote reduce the error to $\delta$.
The equivalent statement follows from $d_{\mathrm H}^2=1-\mathrm{BC}$ by setting $\epsilon=\sqrt\eta$.
\end{proof}
Since a point mass is an admissible reference distribution, this also proves a worst-case lower bound for identity testing with a general known reference distribution. The matching upper bound above assumes that the reference is a point mass.

\subsection{Fidelity to a pure state and \texorpdfstring{$\sqrt{\tr(\rho\sigma^2)}$}{sqrt(tr(rho sigma squared))} estimation}
\label{sec:mixed-fidelity}
Fang and Wang~\cite[Theorems~6 and~8]{FW25} give constant error estimators for $\mathrm F(\rho,\ket{\psi})=\sqrt{\bra{\psi}\rho\ket{\psi}}$ and $\sqrt{\tr(\rho\sigma^2)}$, each using $O(1/\epsilon)$ queries.
Here $\rho$ and $\sigma$ are density operators, and purified quantum query access allows queries to a unitary preparing a purification of each input state, its inverse, and their controlled versions.

\begin{corollary}
\label{cor:mixed-fidelity}
For $0<\epsilon\le1/32$ and $0<\delta\le1/3$, estimating either $\mathrm F(\rho,\ket{\psi})$ or $\sqrt{\tr(\rho\sigma^2)}$ to additive error $\epsilon$ with failure probability at most $\delta$ has purified quantum query complexity $\Theta(\log(1/\delta)/\epsilon)$.
\end{corollary}
\begin{proof}
Run the respective estimator of~\cite{FW25} independently $O(\log(1/\delta))$ times and take the median, giving the upper bound.
For the lower bound, restrict to $\rho=\ketbra{\varphi}{\varphi}$ and $\sigma=\ketbra{\psi}{\psi}$ with trivial purifying registers. Both quantities then equal $|\braket{\varphi}{\psi}|$, so \Cref{thm:fidelity-small-error-lb} applies.
\end{proof}

\begin{ack}
This research was partially supported by the Quantum Science and Technology-National Science and Technology Major Project under Grant No.~2024ZD0300500.
Qisheng Wang was supported by the Startup Funding from Shanghai Jiao Tong University.
\end{ack}

\bibliographystyle{plainnat}
\bibliography{arxiv}

\end{document}